\documentclass{iau}
\usepackage{graphicx}

\title[Secular migration of close-in exoplanets with tidal and magnetic torques] 
{Could star-planet magnetic interactions lead to planet
migration and influence stellar rotation ?}

\author[J\'er\'emy Ahuir, Antoine Strugarek, Allan-Sacha Brun et al.]   
{J\'er\'emy Ahuir$^1$, Antoine Strugarek$^1$, Allan-Sacha Brun$^1$, St\'ephane Mathis$^1$, Emeline Bolmont$^{2,1}$
Mansour Benbakoura$^1$, Victor R\'eville$^{3,1}$ \and Christophe Le Poncin-Lafitte$^4$}

\affiliation{$^1$D\'epartement d'Astrophysique-AIM, CEA/IRFU, CNRS/INSU, Universit\'e Paris-Saclay, Universit\'e Paris Diderot, Universit\'e de Paris, 91191 Gif-sur-Yvette, France \\ email: {\tt jeremy.ahuir@cea.fr} \\[\affilskip]
$^2$Observatoire Astronomique de l'Universit\'e de Gen\`eve, Universit\'e de Gen\`eve, CH-1290 Versoix, Switzerland\\[\affilskip]
$^3$ IRAP, Universit\'e Toulouse III - Paul Sabatier, CNRS, CNES, Toulouse, France \\[\affilskip]
$^4$SYRTE, Observatoire de Paris, Universit\'e PSL, CNRS, Sorbonne Universit\'e, LNE, 61 avenue de l'Observatoire, 75014 Paris, France
}

\pubyear{2019}
\volume{354}  
\pagerange{XXX--YYY}
\jname{Solar and Stellar Magnetic Fields: Origins and Manifestations}
\editors{A. Kosovichev, K. 
G. Strassmeier, M. Jardine, eds.}
\begin{document}

\maketitle

\begin{abstract}
The distribution of hot Jupiters, for which star-planet interactions can be significant, questions the evolution of exosystems. We aim to follow the orbital evolution of a planet along the rotational and structural evolution of the host star by taking into account the coupled
effects of tidal and magnetic torques from \textit{ab initio} prescriptions. It allows us to better understand the evolution of star-planet systems and to explain some properties of the
distribution of observed close-in planets.
To this end we use a numerical model of a coplanar circular star-planet system taking into account stellar structural changes, wind braking and star-planet interactions, called ESPEM (\cite[Benbakoura et al. 2019]{benbakoura}). We find that depending on the initial configuration of the system, magnetic effects can dominate tidal effects during the various phases of the evolution, leading to an important migration of the planet and to significant changes on the rotational evolution of the star. Both kinds of interactions thus have to be taken into account to predict the evolution of compact star-planet systems. 

\keywords{stellar evolution, solar-type stars, stellar rotation, magnetism, star-planet interactions}
\end{abstract}

\firstsection 
              
\vspace*{-0.50cm}
\section{Introduction}

The discovery of more than 4000 exoplanets during the last two decades has shed light on the importance of characterizing star-planet interactions. Indeed, a large fraction of these planets have short orbital periods and are consequently strongly interacting with their host star. In particular, several planetary systems, like 55 Cancri, WASP-18 or HD 189733 (cf. Figure \ref{ESPEM} to see the architecture of those systems), are likely to host exoplanets undergoing a migration due to tidal and magnetic torques. We consider here the joint influence of stellar wind, tidal and magnetic star-planet interactions on the star's rotation rate and planetary orbital evolution. We focus our study on the relative influence of tidal and magnetic torques on the system evolution. Our objective is to take into account simultaneously \textit{ab initio} prescriptions of tidal and magnetic torques in exosystems, so as to improve our understanding of close-in star-planet systems and their long-term evolution. 

\vspace*{-0.30cm}
\section{Stellar rotation and planetary migration}
ESPEM (French acronym for Planetary Systems Evolution and
Magnetism; \cite[Benbakoura et al. 2019]{benbakoura}) is a 1D numerical model of a coplanar circular star-planet system allowing us to follow the joint secular evolution of the semi-major axis of the orbit and the stellar rotation. We consider in this model a solar-type star with a bi-layer internal structure following \cite[MacGregor \& Brenner 1991]{mcgregorbrenner}, whose changes are monitored along the evolution of the system by relying on grids provided by the 1D stellar evolution code STAREVOL (\cite[Amard et al. 2016]{amard}). The companion is considered a punctual mass.
 \begin{figure}[!h]
        \centering
        \includegraphics[scale=0.36]{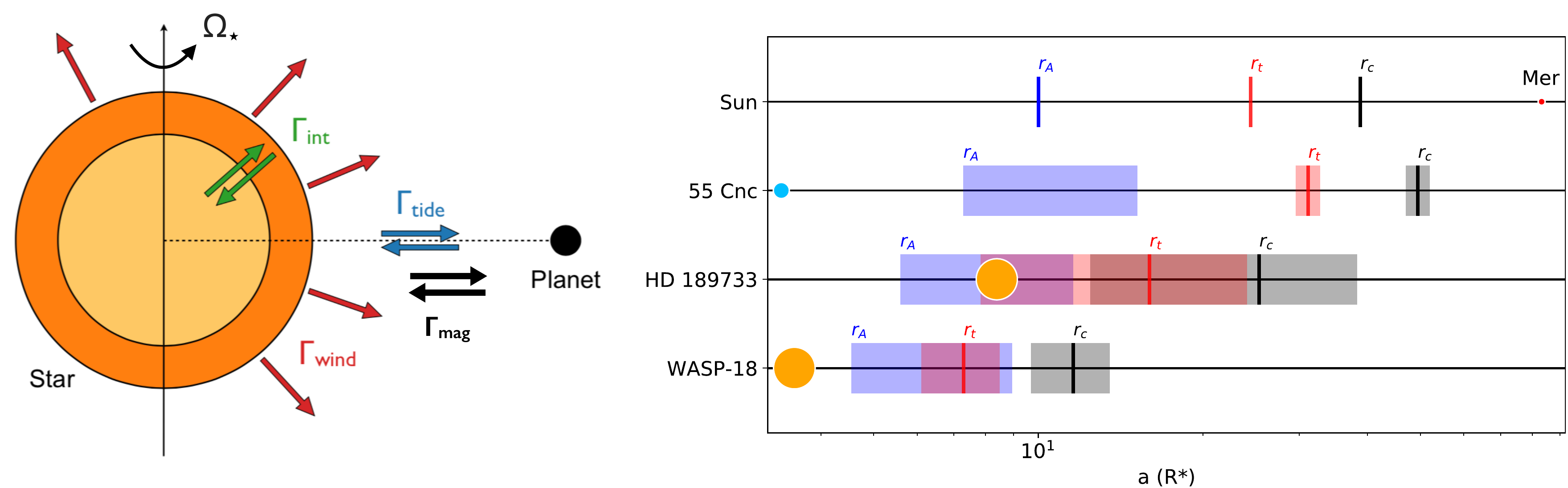}
  \caption{\label{ESPEM} On the left: schematic view of the system and its interactions (adapted from \cite[Benbakoura et al. 2019]{benbakoura}). The various \(\Gamma\) quantities (int, wind, mag, tide) illustrate the various angular momentum transfer mechanisms taken into account in our study. On the right: architecture of the inner Solar System, 55 Cancri, HD 189733 and WASP 18. The semi-major axis \(a\) is expressed in stellar radii \(R_\star\). Red dots: Sub-Earths. Blue dots: Super-Earths. Orange dots: jovian planets. In black: co-rotation radius \(r_c\), with error bars. In red: transition equilibrium-dynamical tide \(r_t\), with error bars. In blue: Estimates of the sub-alfv\'enic region transition \(r_A\) for realistic stellar magnetic fields.}
\end{figure}

A global schematic view of the different interactions involved in the system is given in Figure \ref{ESPEM}. The yellow disk depicts the stellar radiative core and the orange shell the convective envelope. The red arrows account for the angular momentum extraction by the stellar wind (\cite[Schatzman 1962]{schatzman}, \cite[Weber \& Davis 1967]{webdav}; see \cite[Ahuir et al. 2019]{ahuir19} for the interdependencies between magnetism, wind and stellar rotation). The green arrows correspond to internal angular momentum exchanges between the core and the envelope of the star, which in absence of external disruptions evolve towards synchronization of their rotation (\cite[MacGregor \& Brenner 1991]{mcgregorbrenner}). Such a coupling also takes into account the growth of the radiative zone during the Pre-Main Sequence, resulting in an angular momentum transfer from the envelope to the core (\cite[MacGregor 1992]{mcgregor}). 
The blue arrows account for exchanges between the stellar envelope and the planetary orbit due to both equilibrium and dynamical tidal effects (\cite[Hansen 2012]{hansen}, \cite[Ogilvie 2013]{ogilvie13}, \cite[Mathis 2015]{mathis15}, \cite[Bolmont \& Mathis 2016]{bolmontmathis}).

\vspace*{-0.35cm}
\section{Two-body magnetic interaction}
Along with wind braking and tidal effects, magnetic interactions (black arrows in Figure \ref{ESPEM}) occur because of the relative motion between the planet and the ambient wind at the planetary orbit. If the planet is in a sub-alfvenic region (where the wind velocity is smaller than the local Alfv\'en speed), a magnetic torque applied to the planet can be associated to an efficient transport of angular momentum between the planet and the star through the so-called \textit{Alfv\'en wings} (\cite[Neubauer 1998]{neubauer}). Several regimes then appear according to the magnetic properties of the planet, at least as a first approximation. If the planet is able to sustain a magnetosphere, Alfv\'en waves generally do not have enough time to go back and forth between the star and the planet. In this case, two uniques Alfv\'en wings form around the planet, leading to the so-called \textit{dipolar} interaction (\cite[Saur et al. 2012]{saur}, \cite[Strugarek et al. 2015]{strugarek15}, \cite[Strugarek 2016]{strugarek16}). In the opposite case, when the ambient wind pressure is too strong or if the planetary dynamo sustains a too weak magnetic field,  propagating Alfv\'en waves can overlap and the interaction becomes \textit{unipolar} (\cite[Laine et al. 2008]{laine08}, \cite[Laine \& Lin 2012]{lainelin}). \cite[Strugarek et al. (2017)]{strugarek17} have performed a first study on planetary migration taking into account simultaneously tidal and magnetic torques, by
computing the migration timescale of the planet for both contributions. Their study reveals that both effects could play a key role depending on the
characteristics of the star-planet system considered. A self-consistent secular evolution of star-planet systems under the influence of magnetic and tidal torques is
thus needed to properly disentangle the importance of the two physical mechanisms.

\vspace*{-0.27cm}
\section{Influence of the tidal and MHD torques on the orbital evolution and the stellar rotation}

To assess the relative contributions of the tidal and magnetic torques on the star-planet secular evolution, we will assume in the following section that the planet is located in an open-field region to maximize magnetic interactions (for example in the equatorial plane of a star with a quadrupolar magnetic field). The stellar magnetic field is then assumed to be radial
(\(B_\star \sim 1/r^2\)).\\
 \begin{figure}[!h]
        \centering
        \includegraphics[scale=0.36]{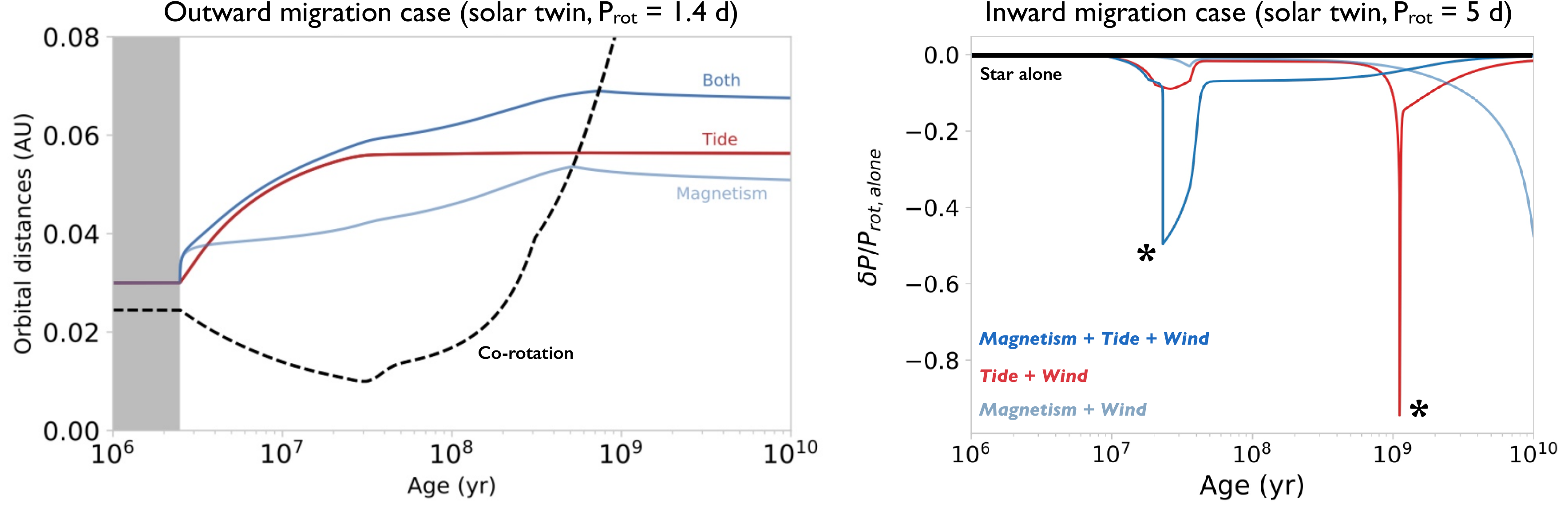}
 \caption{\label{evol} On the left: orbital evolution of a Jupiter-like planet, initially at \(a = 0.03\) AU, orbiting a fast rotating solar twin (\(P_\textrm{rot} = 1.4\) d) in the open-field configuration. The black dashed line corresponds to the co-rotation radius of the star without any planet. On the right: evolution of the relative difference of rotation periods between a midly rotating solar twin (\(P_\textrm{rot} = 5\) d)  hosting a Jupiter-like planet initially at \(a = 0.03\) AU and a similar star without planet. Mark \textbf{*} corresponds to the planetary destruction. In black: case of a star without planet.}
\end{figure}

The left panel of Figure \ref{evol} shows the evolution of the semi-major axis of a Jupiter-like planet, initially at a distance \(a=0.03\) AU, orbiting around a fast rotating solar twin \((P_\textrm{rot} = 1.4\) d). The coupled action of the tidal and magnetic torques (in dark blue in Figure \ref{evol}) shows an outward migration acting at the beginning of the evolution and an evolution of the semi-major axis deviating significantly from the case where tidal effects alone are acting on the system (in red in Figure \ref{evol}). The orbital evolution of the planet taking into account star-planet magnetic interactions without tidal effects (in light blue in Figure \ref{evol}) also presents an important outward migration at the beginning of the simulation, then a less pronounced evolution than in the purely tidal case.\\

The magnetic and tidal torques lead to angular momentum exchange between the orbital motion of the planet and the rotating star. Because of the conservation of the total angular momentum of the system, stellar rotation is also affected by those interactions. The right panel of Figure \ref{evol} shows a case of planet destruction, for which a Jupiter-like planet located initially at \(a = 0.03\) AU orbits a midly rotating solar twin (\(P_\textrm{rot} = 5\) d). The influence of planetary migration on stellar rotation is quantified by
\begin{equation}
\delta P = P_\textrm{rot}-P_\textrm{rot,alone},
\end{equation}
where \(P_\textrm{rot,alone}\) is the rotation period of the star without a planet. In the case of a planetary destruction (mark \textbf{*} in Figure \ref{evol}), the star presents a sharp spin-up when the planet falls down in the central body. This corresponds to the transfer of its orbital angular momentum to the host-star. The presence of magnetic torques (in dark blue in Figure \ref{evol}) changes the survival time of the planet, making its destruction happen earlier than in the purely tidal-case. In the case of Figure \ref{evol} the planet falls during the PMS, which is at the origin of a briefer and less intense spin-up than for the tide-only case (in red in Figure \ref{evol}). 
An evolution only driven by magnetic effects (in light blue in Figure \ref{evol}) leads to a potential destruction of the planet after the end of the simulation, which is at the origin of a monotonous decrease of the stellar rotation rate. In those three configurations, an overestimated stellar rotation rate due to planetary migration, compared to what it is expected from gyrochronology (\cite[Skumanich 1972]{skumanich}, \cite[Barnes 2003]{barnes}), can last several billions of years.

\vspace*{-0.29cm}
\section{Conclusions}
We presented some preliminary results obtained with the ESPEM model comparing the relative importance of tidal and magnetic torques for exosystems' evolution. Depending on the initial configuration of the system, magnetic effects can dominate tidal effects along the evolution of the system, which is in agreement with \cite[Strugarek et al. (2017)]{strugarek17}.
Furthermore, the planet has a significant influence on stellar rotation, especially in the case of an efficient inward migration and a collision during the MS. First statistical studies performed with ESPEM tend to show that initial fast stellar rotation result in the excitation of tidal inertial waves and of a stronger magnetic field, that leads to intense tidal-magnetic effects and therefore to more efficient planetary migration (\cite[Teitler \& Königl 2014]{teitler}). Studying the evolution of a synthetic population of exoplanets may allow us to explain some features of the distribution of close-in planets (\cite[McQuillan et al. 2013]{mcquillan}). We intend to explore such statistical approaches in the near future. With a similar wind prescription as the one used in ESPEM we find that 20\% of the exoplanets detected so far are likely to lie in a sub-alfvenic region. The dissipation of tidal gravity waves (\cite[Zahn 1975]{zahn75}, \cite[Goodman \& Dickson 1998]{goodman}, \cite[Terquem et al. 1998]{terquem}) will be also taken into account in future work.

The authors acknowledge funding from the European
Union’s Horizon-2020 research and innovation programme (Grant Agreements no. 776403 ExoplANETS-A and no. 647383 ERC CoG SPIRE), INSU/PNST, INSU/PNP and the CNES-PLATO grant at CEA. This work has been carried out within the framework of the NCCR PlanetS supported by the Swiss National Science Foundation. This research has made use of NASA's Astrophysics Data System.

\end{document}